\documentclass[a4paper,10pt]{article}

\usepackage[utf8]{inputenc}
\usepackage{fullpage}
\usepackage{amsmath}
\usepackage{amsfonts}
\usepackage{amssymb}
\usepackage[pdftex]{hyperref}
\usepackage{graphicx}
\usepackage{algorithmic}
\usepackage{algorithm}
\usepackage{subfigure}
\usepackage[numbers,sort&compress]{natbib}
\usepackage{authblk}

\title{On the long-term correlations and multifractal properties of electric arc furnace time series}

\author[1]{Lorenzo Livi\thanks{llivi@scs.ryerson.ca}\thanks{Corresponding author}}
\author[2]{Enrico Maiorino\thanks{enrico.maiorino@uniroma1.it}}
\author[2]{Antonello Rizzi\thanks{antonello.rizzi@uniroma1.it}}
\author[1]{Alireza Sadeghian\thanks{asadeghi@ryerson.ca}}
\affil[1]{Dept. of Computer Science, Ryerson University, 350 Victoria Street, Toronto, ON M5B 2K3, Canada}
\affil[2]{Dept. of Information Engineering, Electronics, and Telecommunications, SAPIENZA University of Rome, Via Eudossiana 18, 00184 Rome, Italy}

\providecommand{\keywords}[1]{\textbf{\textit{Keywords---}} #1}

\begin{document}

\maketitle

\begin{abstract}
In this paper, we study long-term correlations and multifractal properties elaborated from time series of three-phase current signals coming from an industrial electric arc furnace plant.
Implicit sinusoidal trends are suitably detected by considering the scaling of the fluctuation functions.
Time series are then filtered via a Fourier-based analysis, removing hence such strong periodicities.
In the filtered time series we detected long-term, positive correlations.
The presence of positive correlations is in agreement with the typical V--I characteristic (hysteresis) of the electric arc furnace, providing thus a sound physical justification for the memory effects found in the current time series.
The multifractal signature is strong enough in the filtered time series to be effectively classified as multifractal.\\
\keywords{Multifractal analysis; Electric arc furnace; Time series analysis; Fourier detrending.}
\end{abstract}

\section{Introduction}

The analysis of complex systems \citep{friedrich2011approaching,kwapien2012physical,crutchfield2012between} is recently gaining traction with the development of new computational and theoretical tools, specifically designed to tackle the problem of disentangling the complex interactions among different parts of a process.
A common approach for this purpose is the study of scale-invariant properties \cite{stanley1999scaling}, i.e., those features that manifest in a similar manner at all magnification levels of the system. The scale-invariance property is a peculiar feature of fractal structures, and it is quantifiable by evaluating a suitable collection of scaling exponents that compactly describe how the system evolves when observed at different scales \cite{harte2010multifractals}.
In particular, the scaling of the correlation properties of a stochastic process could provide valuable information on the underlying dynamical system, especially in cases in which the observed system denotes long-term memory and nonlinear behaviors \cite{kantelhardt2009fractal,drozdz2009quantitative}.

Long-term memory properties of a stochastic process are firstly described by power-law like decreasing of the autocorrelation calculated for suitable observables.
This feature can be characterized by the value of the so-called Hurst exponent $H$ \cite{serinaldi2010use}, which assumes values in $[0, 1]$.
The value of $H$ is 0.5 when the process corresponds to uncorrelated noise, whereas if the process is persistent (correlated) or antipersistent (anticorrelated) then $H$ is greater than or less than 0.5, respectively.
However, conventional methods employed to analyze the long-term correlations of a time series (e.g., spectral analysis and Hurst analysis) are known to be misleading when data are non-stationary and contains deterministic trends \cite{PhysRevE.65.041107,PhysRevE.64.011114}.
Therefore, in many experiments it is important to distinguish intrinsic fluctuations characterizing the process from pure trending behaviors.
A well-established approach employed to this end is the Detrended Fluctuation Analysis (DFA) \cite{shao2012comparing}, which has been successively generalized in the so-called Multifractal Detrended Fluctuation Analysis (MF-DFA) \cite{kantelhardt2002multifractal,PhysRevE.74.016103,kantelhardt2009fractal}. MF-DFA accounts for the possibility to characterize different degrees of correlation by considering a collection of scaling exponents.
Such scaling exponents provide a way to perform classical multifractal analyses on time series.
Notably, it is possible to calculate the so-called multifractal spetrum, which describes all relevant multifractal characteristics of the system under analysis.
MF-DFA technique has been used in a multitude of contexts, such as breast cancer imaging \cite{gerasimova2013multifractal}, river flow modeling \cite{movahed2008fractal}, brain signals \cite{fetterhoff2014multifractal}, solar filaments and sunspots \cite{hu2009multifractal,Wu20151}, financial returns \cite{zhou2009components,meng2012effects}, high-energy physics \cite{mali2015multifractal}, earth's gravity \cite{Telesca2015426}, protein contact networks \cite{mixbionets2}, and protein molecular dynamics \cite{zhou2014fractal}.

Detrending is a very crucial aspect in multifractal analysis of time series \cite{bashan2008comparison,ludescher2011spurious,PhysRevE.71.011104}.
In fact, trends are usually a source of spurious multifractal signatures observed in the analyzed series.
Conventional MF-DFA is able to deal with trends compatible with (higher-order) polynomials.
Nonetheless, trending could appear also as a consequence of periodicities, such as those induced by seasonality for long time series.
To this end, \citet{chianca2005fourier} proposed to perform a preprocessing stage involving a Fourier analysis in order to identify and then remove specific frequencies with large amplitude.
Such a filtering technique demonstrated to provide good results in several different contexts \cite{nagarajan2005minimizing,movahed2008fractal,zhao2011multifractal}, while however criticisms and limitations of the method have been highlighted in the context of sunspot time series \cite{hu2009multifractal}.

In this paper, we study long-term correlations and multifractal properties elaborated from time series of three-phase current signals coming from an electric arc furnace (EAF) \cite{sadeghian1999nonlinear,sadeghian2011dynamic,sadeghian1999application}.
EAF modeling and analysis of related data is an active area of research \cite{1413319,6690260,janabi2009adaptive,samet2014dynamic,samet2012wide,samet2014maximizing}. Previous studies focused mostly on system identification and prediction problems with the aim of power optimization and control.
However, to our knowledge there is no track in the literature of studies employing MF-DFA to characterize EAF signals.
An EAF is a plant utilized in industrial environments for the production of steel by a melting process of metallic scraps.
EAF consists of a cylindrical melting pot covered by a removable vault, inside of which are placed one or more electrodes.
The scraps to be molten (metallic charge) are deposited on the bottom of the melting pot and, by applying AC or DC tension to the electrodes, an electric arc is generated between the extremities of such electrodes and the scraps. The current traversing the metallic charge raises its temperature until the fusion point is reached, initiating the melting process.
The V--I characteristic of the EAF is the result of a nonlinear process with memory \cite{sadeghian2011dynamic}, since the electrical current value is determined as a function of both the voltage and the internal state, which in turn is determined by the past states of the system.
Such a process gives rise to an electrical phenomenon called \textit{hysteresis} \cite{chua1987linear}.

Toward removing all implicit periodicities proper of an alternating current setting, the EAF time series are preprocessed via a Fourier-based analysis in order to remove the first few complex coefficients.
By analyzing the filtered time series we detected long-term, positive correlations. This outcome is in agreement with the typical V--I characteristic of the EAF, which shows a clear hysteretic behavior, as in fact the equivalent dipole of the arc is a dynamic and nonlinear one \cite{chua1987linear}.
The multifractal signature is consistent among the three-phases and it is strong enough in the filtered time series to be effectively classified as multifractal.
Finally, we show that shuffling the series reveals that the multifractal signature is entirely due to the long-term correlations.

The paper is structured as follows.
Section \ref{sec:eaf} provides an introduction to the EAF context.
In Section \ref{sec:results} we discuss our results regarding long-term correlations and multifractal properties of the analyzed three-phase current signals.
Section \ref{sec:conclusions} provides the concluding remarks and future directions.
Finally, in Appendix \ref{sec:mfdfa} we introduce the MF-DFA technique.

\section{Electric arc furnace and the V--I characteristic}
\label{sec:eaf}

The analyzed EAF \cite{sadeghian2011dynamic} is an alternating current furnace powered by a three-phase electrical supply. 
The plant contains three electrodes placed on the three vertices of a triangle, supported by vertical guides that allow controlling their distance with respect to the underlying metallic charge.
When the applied voltage is sufficiently high to break the air resistance, an electric arc is struck from the extremity of each electrode to the metallic charge.
In an ideal setting, the electric arc is created each time the absolute value of the alternating voltage exceeds a threshold and, consequently, the current assumes an almost sinusoidal waveform.
However, in real-world situations, the V--I characteristic is subject to many sources of noise, such as arc instabilities, non-monochromaticity of the input signal, and metallic charge cave-ins that hinder the analysis of the measured data.
The latter effect, for example, might cause sudden drops in the current magnitude.

\begin{figure}[h]
\centering
\includegraphics[viewport=0 0 654 456,scale=0.30,keepaspectratio]{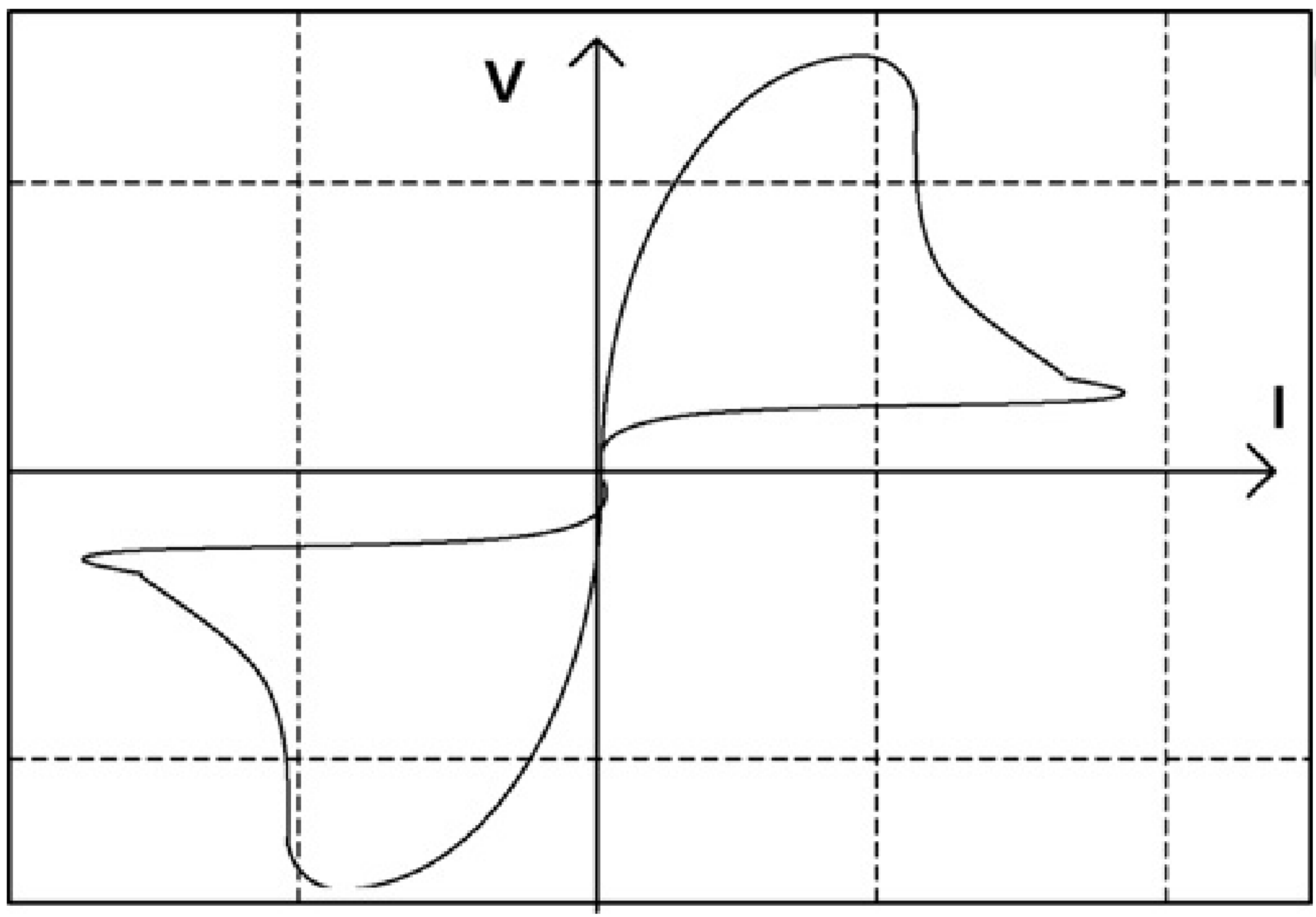}
\caption{Hysteresis of the V--I characteristic of EAF. Image courtesy of \citet{sadeghian2011dynamic}.}
\label{fig:hysteresis}
\end{figure}

The V--I characteristic of the system forms a closed trajectory in the tension--current phase space (see Fig. \ref{fig:hysteresis}). This behavior denotes a branching process in the joint evolution of those two quantities and it is the distinctive feature of a system subject to hysteresis \cite{chua1987linear}.
This implies that the system possesses memory of the previous states.
As we will show in this study, the hysteresis can be investigated through the analysis of the long-term correlations of the related EAF current time series.

\section{Results and discussion}
\label{sec:results}

Our data consists in three current time series, each related to one of the three phases of the EAF under analysis \cite{sadeghian2011dynamic}.
In the following, we will denote the three phases as A, B, and C.
Each time series is composed 1150000 samples, having a sampling frequency of 1920 sample per second. The inter-sample time is hence $520.833\times 10^{-6}$ seconds, which gives a total of roughly $10$ minutes of recording. All data are standardized prior to processing.

\begin{figure}[ht!]
\centering

\subfigure[Autocorrelation function.]{
\includegraphics[viewport=0 0 349 243,scale=0.6,keepaspectratio=true]{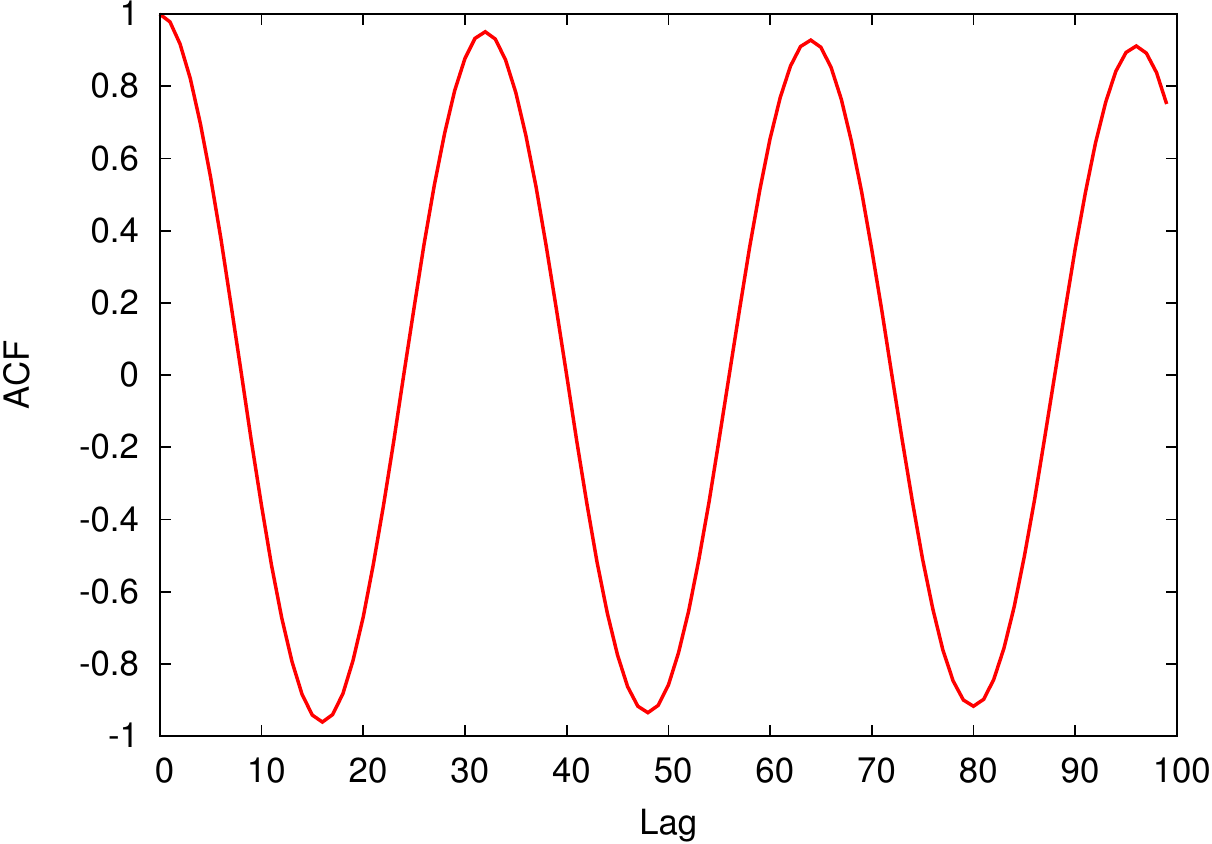}
\label{fig:IA__autocorr}}
~
\subfigure[Power spectrum.]{
\includegraphics[viewport=0 0 343 246,scale=0.6,keepaspectratio=true]{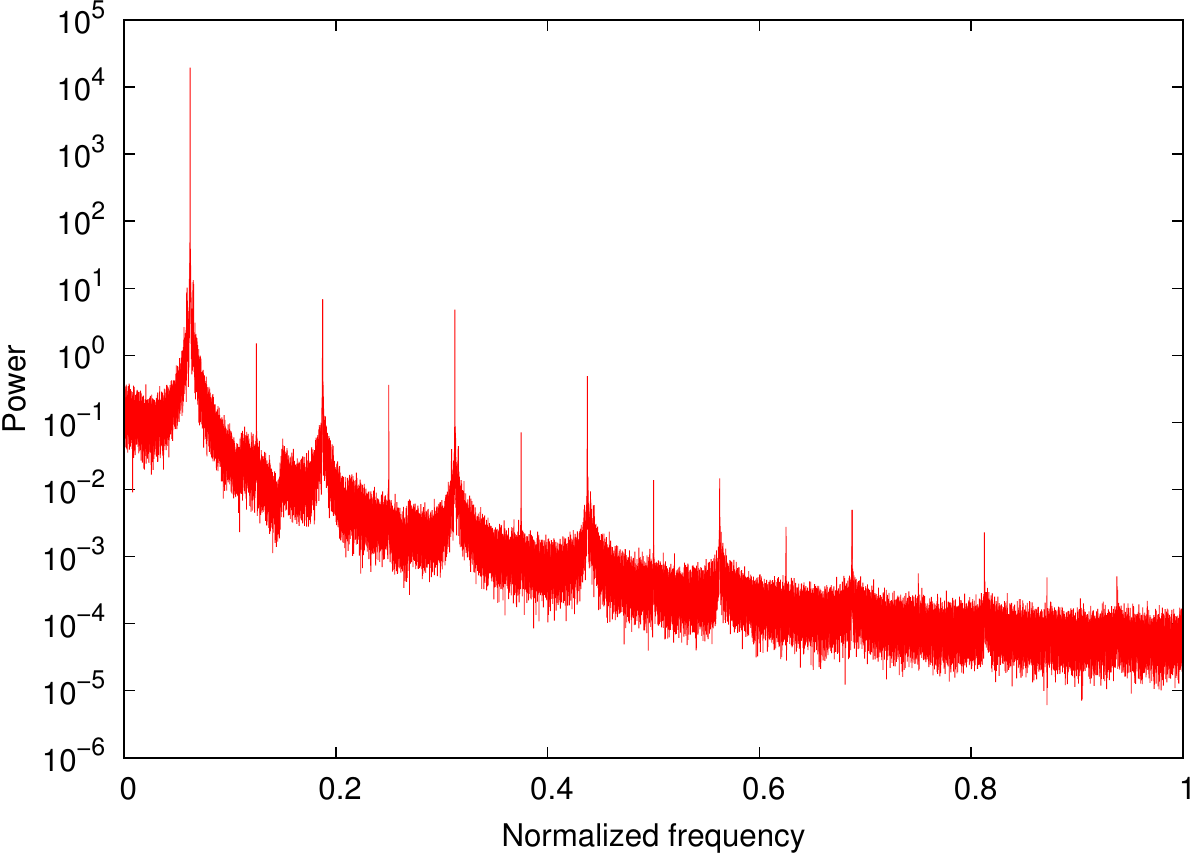}
\label{fig:IA__power}}

\caption{Autocorrelation \ref{fig:IA__autocorr} and power spectrum \ref{fig:IA__power} of current time series of phase A. It is possible to recognize the presence of several harmonics with significant power.}
\label{fig:correlations}
\end{figure}

Fig. \ref{fig:correlations} shows the autocorrelation \ref{fig:IA__autocorr} and power spectrum \ref{fig:IA__power} for phase A.
Autocorrelation is in agreement with a typical sinusoidal waveform of a current signal (60 Hz).
The power spectrum offers an insight on the fact that such a signal contains many harmonics of significant power, which correspond to strong periodicities in the series.
This is fairly expected, due to the nature of the physical system under analysis.
However, in order to avoid the detection of spurious multifractal signatures, it is necessary to eliminate such periodicities.
The interest in the analysis of such EAF signals lies hence in the non-linearities that are super-imposed during the melting process on the baseline sinusoidal signal.

\begin{figure}
 \centering
 \includegraphics[viewport=0 0 344 244,scale=0.6,keepaspectratio=true]{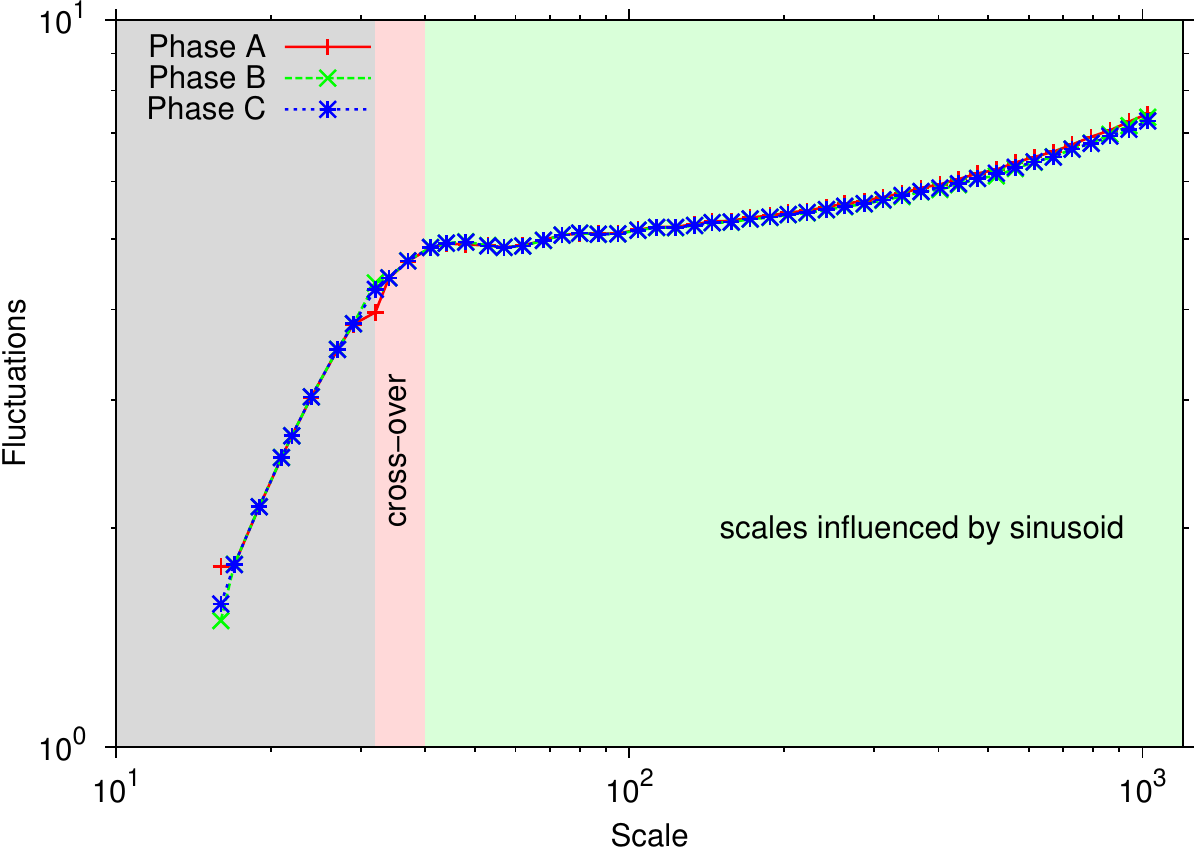}
 \caption{(colors online) Scaling of fluctuations of original (standardized) time series. Fluctuations, calculated with a linear fitting and $q=2$, denote a cross-over compatible with the 60 Hz of the electrical current.}
 \label{fig:fluctuations_crossover}
\end{figure}

\begin{figure}[ht!]
 \centering
 \includegraphics[viewport=0 0 360 246,scale=1,keepaspectratio=true]{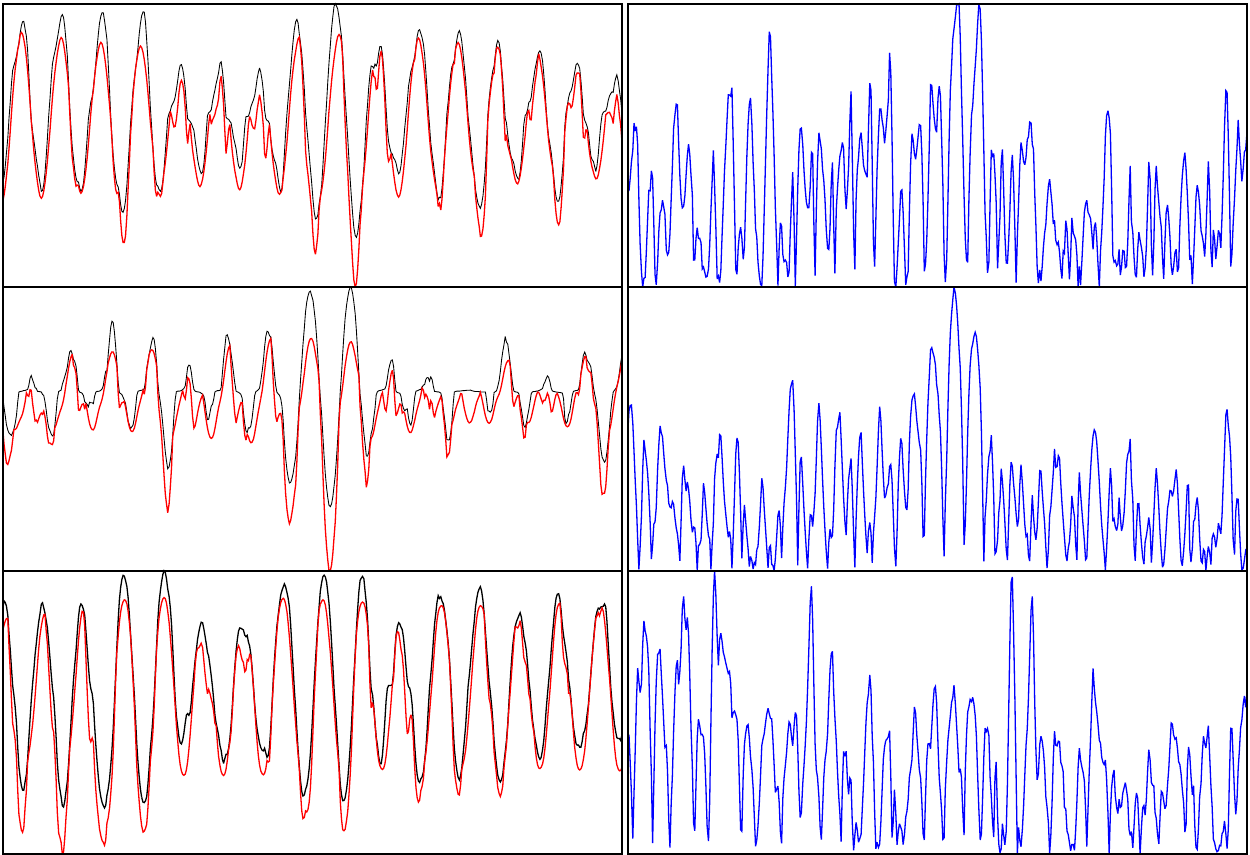}
\caption{(colors online) First 500 samples of the time series (left panels, black lines indicate the original signal while in red we show the trend) and related filtered version (right panels, blue lines). Phase A, B, and C shown from top to bottom.}
\label{fig:signals}
\end{figure}

In Fig. \ref{fig:fluctuations_crossover} we show the scaling of the fluctuations (\ref{eq:Fqshq}) for the three original signals.
A cross-over is visible in Fig. \ref{fig:fluctuations_crossover} at a scale of roughly 32--40 time instants, which corresponds (considering the aforementioned inter-sample time) to a regime of 48--60 Hz. The cross-over is consistent among the three phases, as in fact in the figure the three fluctuation functions overlap almost perfectly.
Therefore, fluctuations, when calculated on the original signals, immediately pick up the main sinusoid of the signal due to the frequency of the electrical current.
In order to remove traces of all periods in the three current signals, we applied Fourier detrending \cite{nagarajan2005minimizing,chianca2005fourier,movahed2008fractal,zhao2011multifractal}. The procedure consists in transforming the signal in the Fourier space with the fast Fourier transform. Complex-valued Fourier coefficients are then order according to their amplitude (from larger to smaller).
In order to obtain a suitable signal to be processed with MF-DFA, we have zeroed the first 3000 coefficients (corresponding to roughly 1.5 seconds) and then antitransformed the signal back to the time domain.
The result of the filtering can be appreciated in Fig. \ref{fig:signals}. In the figure, we show an excerpt of the original signals together with the identified trends (left panels), while in the right panels we report the corresponding filtered versions. It is worth noting that, as expected, the identified trend is not compatible with a single sinusoid, as in fact the trend in the signal is a combination of the main sinusoid with several harmonics of different power (see Fig. \ref{fig:IA__power}).
As we will demonstrate in the following, such filtered signals contain long-term, positive correlations and a fair multifractal signature.

\begin{figure}[ht!]
\centering

\subfigure[Scaling of fluctuations.]{
\includegraphics[viewport=0 0 348 244,scale=0.6,keepaspectratio=true]{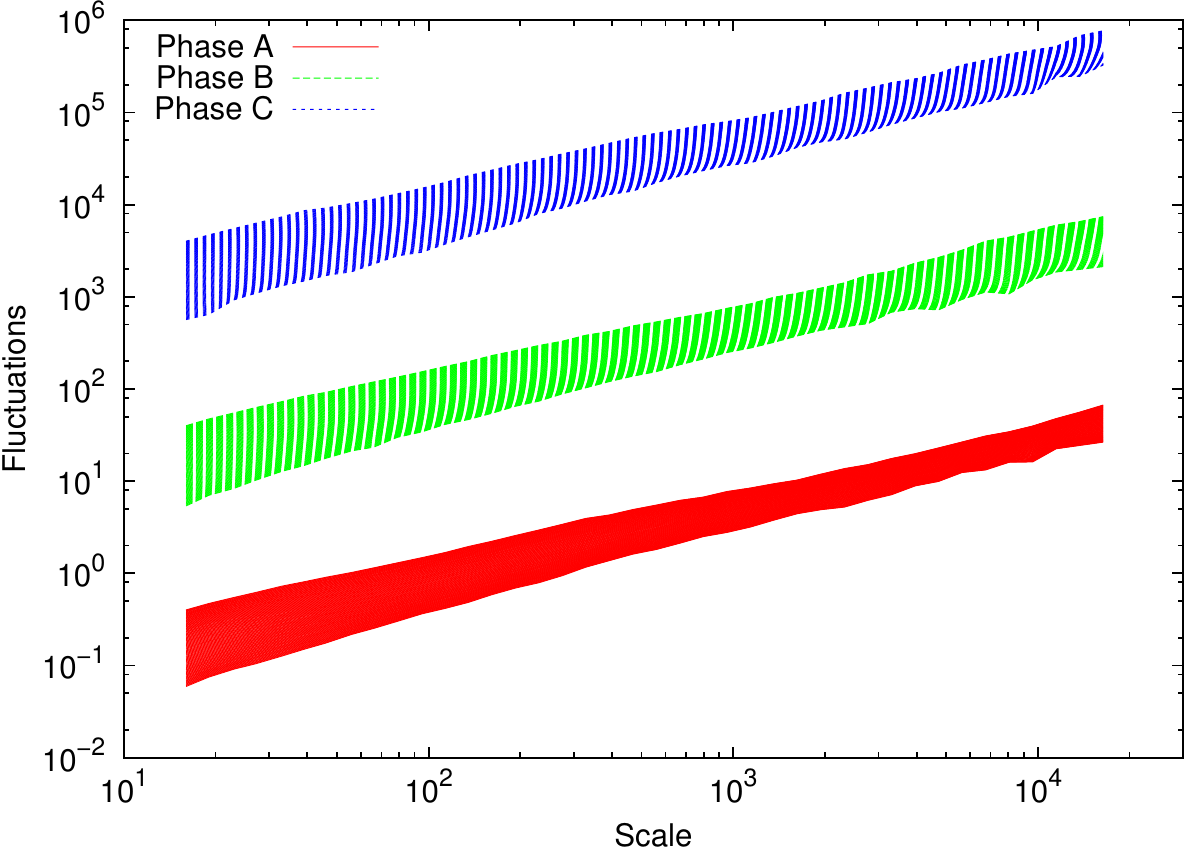}
\label{fig:fluctuations}}
~
\subfigure[Generalized Hurst exponent.]{
\includegraphics[viewport=0 0 345 244,scale=0.6,keepaspectratio=true]{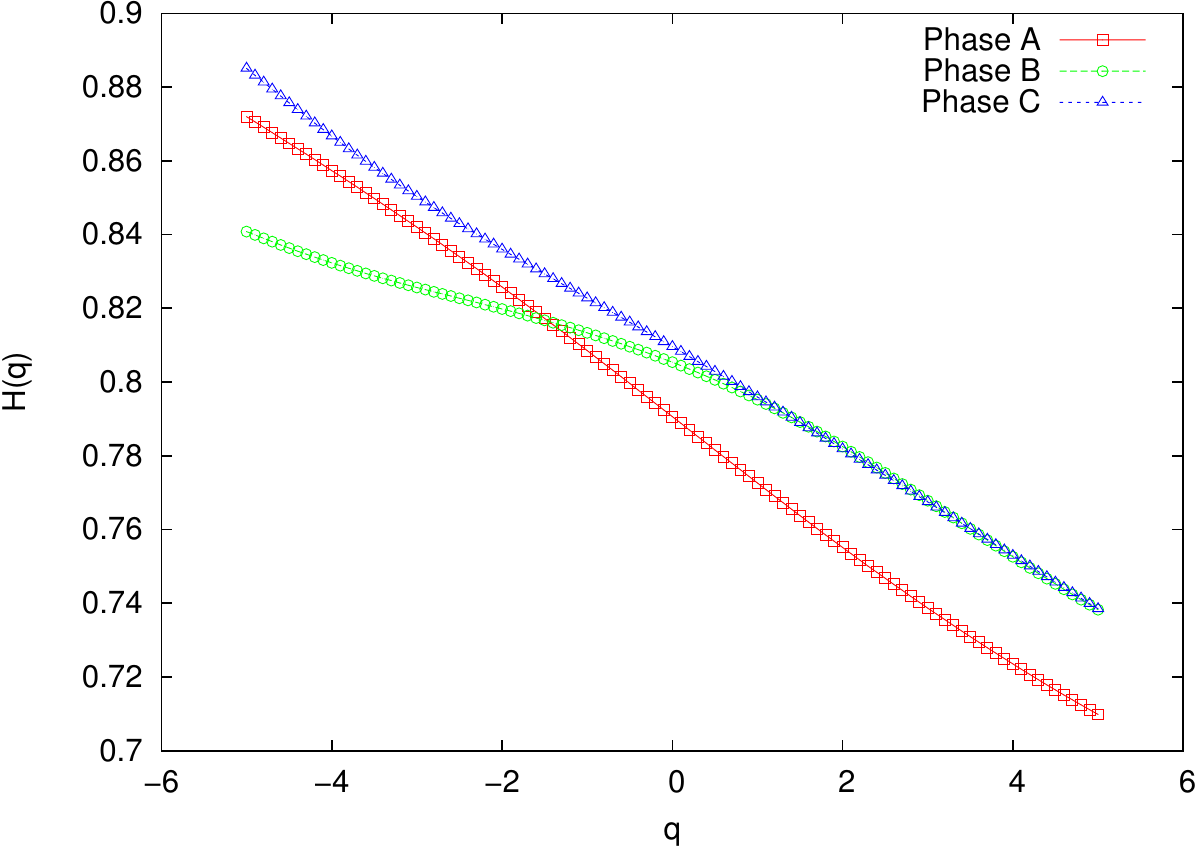}
\label{fig:hurst}}

\subfigure[Local slopes of fluctuations for $q=2$.]{
\includegraphics[viewport=0 0 348 242,scale=0.6,keepaspectratio=true]{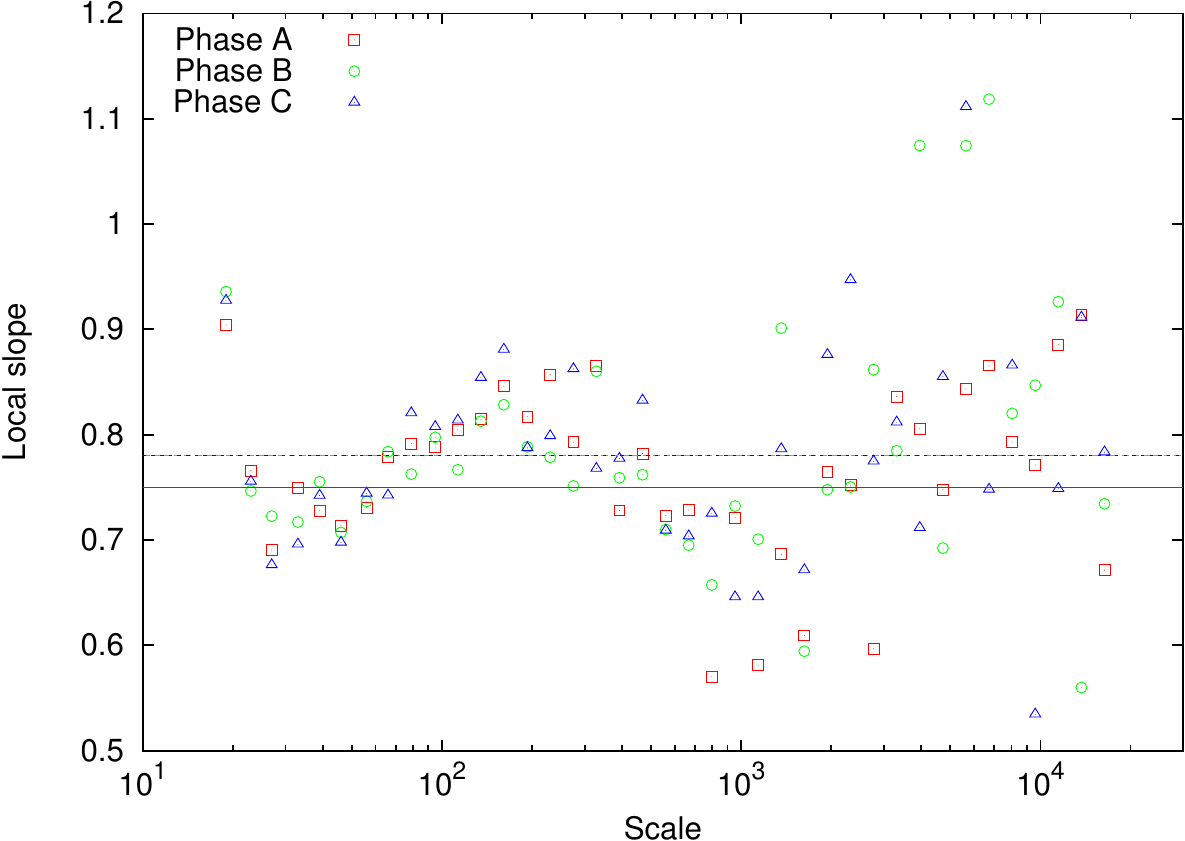}
\label{fig:localslopes}}
~
\subfigure[Multifractal spectrum.]{
\includegraphics[viewport=0 0 343 242,scale=0.6,keepaspectratio=true]{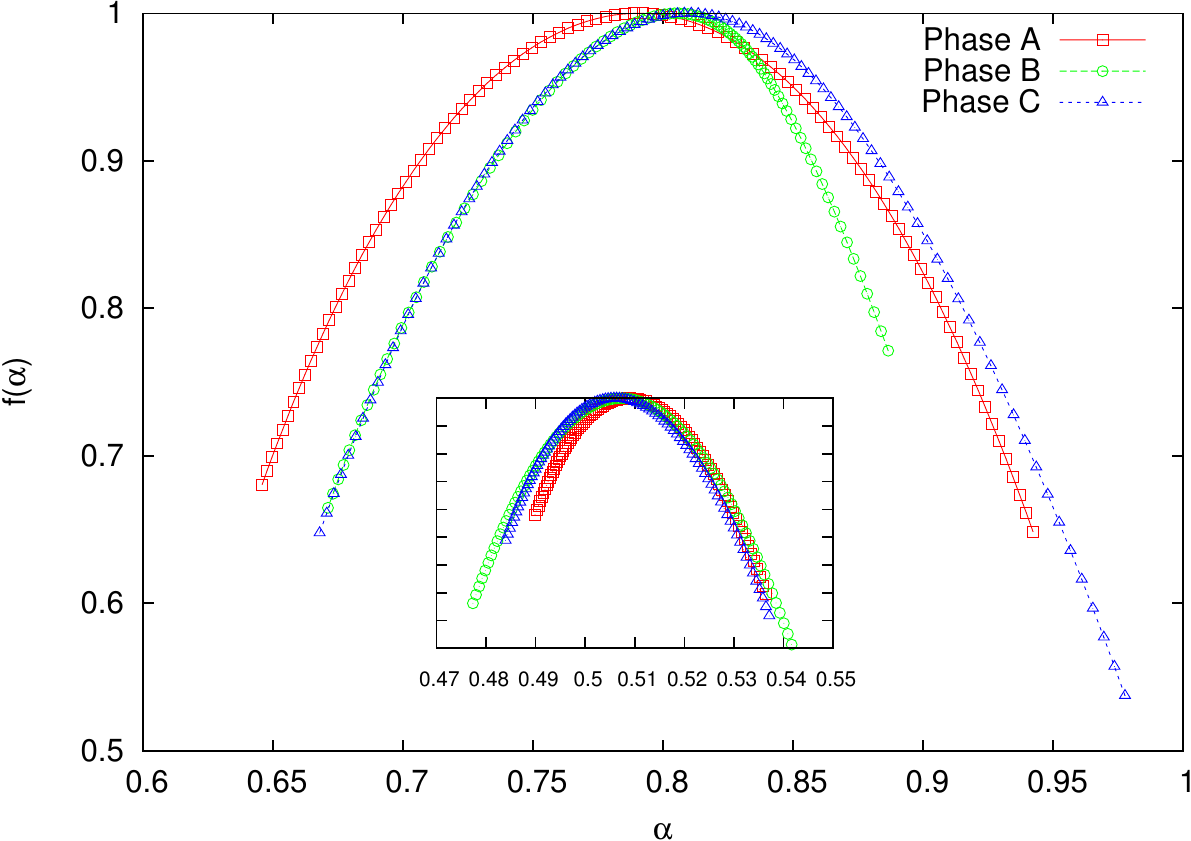}
\label{fig:mfspectra}}

\caption{Scaling of fluctuations \ref{fig:fluctuations}, generalized Hurst exponents \ref{fig:hurst}, local slopes \ref{fig:localslopes} of the fluctuation functions for $q=2$, and multifractal spectra \ref{fig:mfspectra} of the filtered time series. In Fig. \ref{fig:fluctuations}, fluctuation functions are shown for all considered $q$ and vertically shifted to improve visualization. Horizontal lines in Fig. \ref{fig:localslopes} denote the respective estimated Hurst exponents of each phase. The inset in Fig. \ref{fig:mfspectra} shows the spectra obtained after shuffling the time series.}
\label{fig:mfdfa}
\end{figure}

Filtered signals are then analyzed with the MF-DFA procedure, which is described in Appendix \ref{sec:mfdfa}.
MF-DFA has been executed by using scales from 16 up to 16384, $q$ ranging in [-5, 5], and with a linear fitting for the local detrending; same outcome has been obtained with a quadratic fitting. Results are reported in Fig. \ref{fig:mfdfa}.
Scaling of fluctuations, shown in Fig. \ref{fig:fluctuations}, provides a confirmation that the filtering stage removed all relevant periodicities in the original signals. In fact, log-log plot of the filtered signals, for all three phases, is now largely consistent with a global power-law scaling.
Let us now focus on Fig. \ref{fig:hurst}, which shows the generalized Hurst exponents (\ref{eq:Fqshq}).
All three signals denote long-term, positive correlations: $H(2)\simeq 0.75$ for phase A, $H(2)\simeq 0.78$ for phase B and C.
In Fig. \ref{fig:localslopes} we show the local slopes of the fluctuation functions for $q=2$. Local slopes of the three signals fluctuate around the estimated Hurst exponents (the global slope of the entire fluctuation function). As typically found in synthetic time series \cite{shao2012comparing}, higher deviations are observed when moving to larger scales.
The discovery of long-term, positive correlations in all three (filtered) electrical current signals is well-justified and it is in agreement with the typical V--I characteristic of EAF (Sec. \ref{sec:eaf}).
In fact, hysteresis is a phenomenon that is observed when the output of a (nonlinear) system does not depend only on its current input, but also on past system states.
Such a memory effect is imprinted in the EAF time series, as highlighted by the generalized Hurst exponents shown in Fig. \ref{fig:hurst}.

Another important fact in multifractal analysis is the distinction between monofractal and multifractal scaling: if the scaling of a system can be fully characterized by a single exponent (Hurst exponent in our case) then it is monofractal, otherwise it is called multifractal.
The analysis of the multifractal spectrum (\ref{eq:mutifractal_spectrum}) shown in Fig. \ref{fig:mfspectra} provides the answer to such a question.
All three signals denote a fair multifractal scaling: the width (\ref{eq:width}) of the spectrum is $\Delta\alpha \simeq 0.30$ for phase A, $\Delta\alpha \simeq 0.21$ for phase B, and $\Delta\alpha\simeq 0.31$ for phase C. 
It is worth mentioning for phase A and C a small but yet visible tendency of being insensitive to larger fluctuations, which can be deduced from the fact that the spectra are slightly right sided. The converse holds for phase B, which instead appear as slightly left sided -- as in fact the corresponding Hurst exponents tends to be flat for negative $q$.
As discussed by \citet{drozdz2015detecting}, right sided spectra are a less usual in experimental time series.
The authors \cite{drozdz2015detecting} offer a mechanism for reproducing such an effect by constructing a synthetic time series starting from a binomial cascade and successively super-imposing a white noise signal considering a suitable threshold for editing the original cascade.
However, we believe that in our case such an asymmetry is likely to be a byproduct of the preliminary filtering stage performed via Fourier coefficient truncation.
Our results indicate that all three (filtered) time series are compatible with signals having both long-term, positive correlations and a fair amount of multifractal features.
This suggests that the non-linearities and memory effects of the melting process are super-imposed on the original current signals as noise.
Such a noise retains relevant information regarding the physical process, which we suitably detected by means of MF-DFA.

Multifractality could be observed as a byproduct of the long-range correlations and/or it could emerge as a consequence of an underlying broad probability distribution \cite{kantelhardt2009fractal}. Shuffling the time series provides hence a reliable proof to disambiguate the origins of the observed multifractality. In fact, if the multifractal behavior is entirely due to the long-term correlations, then the shuffled signals would not present any multifractality at all; shuffling the data destroys all long-term correlations. This can be quantitatively verified by observing a relevant shrinking of the multifractal spectrum width.
On the other hand, if multifractality is (also) due to a broad underlying density, then shuffling would have only a visible but yet limited impact on the width of the resulting multifractal spectrum.
As shown in the inset of Fig. \ref{fig:mfspectra}, the shuffled signals do not contain any sign of multifractal signature.
This leads us to claim that the observed multifractal properties of the current time series are entirely due to the long-term correlations -- the memory effects due to the hysteresis. We argue that a broad probability density function, in our case, might not be possible, due to the EAF physical constraints. In fact, EAF operates in an industrial setting that is suitably controlled in order to achieve specific operating conditions.
To summarize, in Table \ref{tab:values} we report all relevant coefficients determined via MF-DFA.
\begin{table}[thp!]
\begin{center}
\caption{Summary of all relevant exponents and coefficients found with MF-DFA.}
\begin{tabular}{|c|c|c|c|}
\hline
 & \textbf{phase A} & \textbf{phase B} & \textbf{phase C} \\
\hline
$H(2)$ & 0.75 & 0.78 & 0.78 \\
$\tau(2)$ & 0.51 & 0.56 & 0.56 \\
$\Delta\alpha$ & 0.30 & 0.21 & 0.31 \\
$\Delta\alpha$ (shuffled) & 0.04 & 0.06 & 0.05 \\
\hline
\end{tabular}
\end{center}
\label{tab:values}
\end{table}

\section{Conclusions and future directions}
\label{sec:conclusions}

In this paper, we studied the long-term correlations and multifractal properties elaborated from time series of three-phase electric current measured from an electric arc furnace.
Electric arc furnaces are industrial plants used for the production of steel by a melting process of metallic scraps.
The V--I characteristic of the EAF under analysis is the result of a nonlinear process having memory.
In fact, the value of the electrical current is determined as a function of both voltage and internal state, which in turn is determined by considering the history of the states of the system. Such a process gives rise to an electrical phenomenon called hysteresis.

We initially filtered the original signals via a Fourier based detrending, which we have performed in order to remove the implicit sinusoidal trends due to the alternate current.
Our analysis showed that the memory effects of EAF process are suitably encoded into the filtered time series in the form of long-terms, positive correlations, which are quantified by the high values found for the Hurst exponent.
Successively, we have shown the presence of a fair multifractal signature in all three time series, confirming thus the complexity of such noisy signals.
Finally, shuffling of the (filtered) time series demonstrated that the multifractal signature was entirely due to the long-term correlations of the signals.

Future research works will focus on the study of the long-term and multifractal properties of the cross-correlations between current and voltage time series related to the electric arc furnace.

\appendix{Multifractal detrended fluctuation analysis}
%\section{}
\label{sec:mfdfa}

The MF-DFA procedure is extensively described in Ref. \cite{kantelhardt2002multifractal}.
In this work, we partially relied on the software provided by \citet{ihlen2012introduction}.
The method consists of five steps, three of which are identical to the DFA version.
Given a time series $x_k$ of length $N$ with compact support, the MF-DFA procedure is performed with the following steps:
\begin{itemize}
\item{\textit{Step 1} : The series $Y(i)$ is computed as the cumulative sum (profile) of the series $x_k$:
\begin{equation}
Y(i) = \sum_{k=1}^i \left[ x_k - \langle x \rangle \right], \;\; i = 1, \dotsc, N.
\end{equation}
}
\item{\textit{Step 2} :
The series $Y(i)$ is divided in $N_s \equiv \text{int}(N/s)$ non-overlapping segments of equal
length $s$. 
To account for the possible non-divisibility of $N$ by $s$, the operation is repeated in reverse order by starting from the opposite end of the series, thus obtaining a total of $2N_s$ segments.
}
\item{\textit{Step 3} : The local detrending operation is executed by performing a suitable polynomial fitting on
each of the $2N_s$ segments. Then the variance is determined as
\begin{equation}
F^2(\nu,s) = \frac{1}{s} \sum_{i=1}^s \bigg\{ Y[(\nu-1)s+i] - y_\nu(i) \bigg\}^2,
\end{equation}
for each segment $\nu = 1,\dotsc,N_s$ and
\begin{equation}
F^2(\nu,s) = \sum_{i=1}^s \bigg\{ Y[N-(\nu - N_s)s+i] - y_\nu(i)\bigg\}^2
\end{equation}
for $\nu = N_s +1,\dotsc, 2N_s$, where $y_\nu(i)$ is the fitted polynomial in segment $\nu$. The order $m$ of the fitting polynomial, $y_\nu(i)$, determines the capability of the (MF-)DFA in eliminating trends in the series, thus it has to be tuned according to the expected maximum trending order of the time series.
}
\item{\textit{Step 4} : The $q$th-order average of the variance over all segments is evaluated as
\begin{equation}
\label{eq:Fq}
F_q(s) = \bigg\{ \frac{1}{2N_s} \sum_{\nu=1}^{2N_s} \left[ F^2(\nu,s)\right]^{q/2} \bigg\}^{1/q},
\end{equation}
with $q \in \mathbb{R}$. The $q$-dependence of the fluctuations function $F_q(s)$ allows to highlight the contributions of fluctuations at different magnitude orders.
For $q > 0$ only the larger fluctuations contribute mostly to the average in Eq.~\ref{eq:Fq}; conversely, for $q < 0$ the magnitude of the smaller fluctuations is enhanced. For $q = 2$ the standard DFA procedure is obtained. The case $q = 0$ cannot be computed with the averaging form in Eq.~\ref{eq:Fq} and so a logarithmic form has to be employed,
\begin{equation}
F_0(s) = \exp \bigg\{ \frac{1}{2N_s} \sum_{\nu=1}^{2N_s} \ln \left[F^2(\nu,s)\right] \bigg\}.
\end{equation}
The steps 2 to 4 have to be repeated for different time scales $s$, where all values of $s$ have to be chosen such that $s \geq m+2$ to allow for a meaningful fitting of data.
}
\item{\textit{Step 5} : The scaling behaviour of the fluctuation functions can be determined by analyzing log-log plots of $F_q(s)$ versus $s$ for each value of $q$. If the series $x_i$ is long-range power-law correlated, $F_q(s)$ is approximated (for large values of $s$) by the form
\begin{equation}
\label{eq:Fqshq}
F_q(s) \sim s^{H(q)}.
\end{equation}} 
\end{itemize}

The exponent $H(q)$ is the generalized Hurst exponent; for $q=2$ it reduces to the standard Hurst exponent, $H$. When the considered time series is monofractal, i.e., it shows a uniform scaling over all magnitude scales of the fluctuations, $H(q)$ is independent of $q$. On the contrary, when the small fluctuations scale differently from the large ones, the dependency of $H(q)$ on $q$ becomes apparent and the series can be considered multifractal.

Starting from Eq.~\ref{eq:Fq} and using Eq.~\ref{eq:Fqshq}, it is straightforward to obtain
\begin{equation}
\label{mass_exponent}
Z_q(s)=\sum_{\nu=1}^{N/s} [ F(\nu,s)]^q \sim s^{qH(q) - 1},
\end{equation}
where, for simplicity, it has been assumed that the length $N$ of the series is a multiple of the scale $s$, such that $N_s = N/s$.
The exponent
\begin{equation}
\label{eq:tauq}
\tau(q) = qH(q) - 1
\end{equation}
is the $q$-order mass exponent (also called R\'{e}nyi scaling exponent) of the generalized partition function, $Z_q(s)$. It is worth mentioning that $\tau(2)$ gives an estimation of the correlation dimension.
The multifractal spectrum, denoted as $f(\cdot)$, provides a compact characterization of the multifractality of the time series. Such a function is obtained via the Legendre transform of $\tau(q)$,
\begin{equation}
\label{eq:mutifractal_spectrum}
f(\alpha) = q\alpha - \tau(q),
\end{equation}
where $\alpha$, called singularity exponent, is equal to the derivative $\tau'(q)$. Using Eq.~\ref{eq:tauq} it is possible to directly relate $\alpha$ and $f(\alpha)$ to $H(q)$, obtaining:
\begin{equation}
\alpha = H(q) + qH'(q) \;\; \text{and} \;\; f(\alpha) = q[\alpha - H(q)] + 1.
\end{equation}

The multifractal spectrum (\ref{eq:mutifractal_spectrum}) allows to infer important information regarding the degree of multifractality and the specific sensitivity of the time series to fluctuations of high/low magnitudes.
In fact, the width of the support of $f(\cdot)$, defined as
\begin{equation}
\label{eq:width}
\Delta\alpha=\alpha(q_{-})-\alpha(q_{+}),
\end{equation}
is a direct and important quantitative indicator of the multifractal signature in the series.

\bibliographystyle{abbrvnat}
\bibliography{/home/lorenzo/University/Research/Publications/Bibliography.bib}
\end{document}